# A Mechanical Mass Sensor with Yoctogram Resolution

J. Chaste[1], A. Eichler[1], J. Moser[1], G. Ceballos[1], R. Rurali[2], A. Bachtold[1,*]

(1) Catalan Institute of Nanotechnology, CIN2(ICN-CSIC), Campus de la UAB, 08193 Bellaterra (Barcelona), Spain

(2) Institut de Ciència de Materials de Barcelona (ICMAB-CSIC), Campus de la UAB, 08193 Bellaterra (Barcelona), Spain

**Nanoelectromechanical systems (NEMS) have generated considerable interest as inertial mass sensors. NEMS resonators have been used to weigh cells, biomolecules, and gas molecules, creating many new possibilities for biological and chemical analysis [1-4]. Recently, NEMS-based mass sensors have been employed as a new tool in surface science in order to study e.g. the phase transitions or the diffusion of adsorbed atoms on nanoscale objects [5-7]. A key point in all these experiments is the ability to resolve small masses. Here we report on mass sensing experiments with a resolution of 1.7 yg (1 yg=$10^{-24}$ g), which corresponds to the mass of one proton, or one hydrogen atom. The resonator is made of a ~150 nm long carbon nanotube resonator vibrating at nearly 2 GHz. The unprecedented level of sensitivity allows us to detect adsorption events of naphthalene molecules ($C_{10}H_8$) and to measure the binding energy of a Xe atom on the nanotube surface (131 meV). These ultrasensitive nanotube resonators offer new opportunities for mass spectrometry, magnetometry, and adsorption experiments.**

Mass sensing is based on monitoring the change in the resonance frequency that a resonator experiences when additional mass is adsorbed on its surface. Theoretical analysis predicts that achieving a sensitivity of one yg should be feasible [8,9]. The prospect of reaching this sensitivity level has been motivating tremendous effort for years, because it corresponds to the mass of one proton (1.67 yg) and thus opens the possibility to distinguish between different chemical elements in future inertial mass spectrometry measurements. However, the mass resolution achieved thus far is 7000 yg using microfabricated resonators [10] and as low as about 200 yg using carbon nanotube resonators [11-13] (see Ref. 2).

We combined within a single experiment all the ingredients known to favor the detection of small masses, such as the use of a short nanotube, ultra-high vacuum, and low-noise motion detection. The device consists of a nanotube suspended over a ~150 nm wide trench (Fig. 1a,b). Such a short nanotube produces a sizeable change in the resonance frequency even when a tiny amount of atoms is deposited on the nanotube. Fabrication proceeded by growing nanotubes by chemical vapor deposition on a highly resistive silicon wafer coated with a 1 μm thick oxide layer. Contact electrodes and the side-gate electrode were patterned using electron-beam lithography and Cr/Au evaporation. The trench was etched with fluoridic acid using a PMMA mask (section B of supplementary information). The motion of the nanotube was driven and detected using the frequency modulation (FM) mixing technique at liquid helium temperature, a method known to transduce the nanotube motion into a low-noise electrical signal [14,15]. Figures 1c,d show that the resonance frequency is very high (nearly 2 GHz) and can be tuned with a voltage $V_g$ applied to the gate [16-18]. We developed a computer-controlled feedback loop to monitor the resonance frequency with a feedback time as low as 15 ms (section A of supplementary information). The measurement setup was prepared by lowering the base pressure down to ~3·10$^{-11}$ mbar in order to minimize adsorption of unwanted molecules.

Key to improving mass sensitivity is to anneal the nanotube by passing a large current through it. Prior to annealing, the resonance frequency fluctuates between multiple levels (Fig. 1e, bottom trace); this behavior, observed in all of our resonators, has not been reported thus far, probably because these fluctuations are very small. We found that annealing the nanotube with a current of ~8 µA during ~300 s allows to dramatically reduce these fluctuations (Fig. 1e, top trace). We attribute the origin of these multiple-level fluctuations to a few contamination molecules diffusing on the nanotube surface between various trapping sites (see detailed discussion in section E of supplementary information). Passing a large current removes these molecules. Another advantage of current annealing is that it brings the nanotube back to its initial state after each mass sensing experiment: we continuously tested the device in Fig. 1 during 9 months and observed minor sensitivity degradation over time. This current-induced cleaning process is possible because nanotubes are mechanically robust and chemically inert so they can sustain the large current necessary for cleaning. Resonators made from other materials are usually damaged at such large current densities.

The current annealing cleaning process allows us to achieve an unprecedented mass resolution of 1.7 (±0.5) yg. The inset of Fig. 2 shows that the fluctuations in $f_0$ are remarkably low. The standard deviation $\delta f_0$ is a function of the averaging time $\tau$,

$$\delta f_0 = \left[ \frac{1}{N-1} \cdot \sum_{i=1}^{N} (<f_i^\tau> - <f_0>)^2 \right]^{1/2}$$ with $<f_i^\tau>$ the resonance frequency averaged over

the time interval $i$ with duration $\tau$ and $<f_0>$ the resonance frequency averaged over the whole measurement (Fig. 2). Using $\delta m = 2 m_{eff} \delta f_0 / f_0$ along with the effective mass $m_{eff}$ = 3 (±0.8) · $10^{-19}$ g (estimated from the length and the diameter of the nanotube; section C of supplementary information), we obtain a mass resolution of 1.7 (±0.5) yg after 2 s averaging time. This ultra-low value corresponds to about the mass of one proton (1.67 yg). The error in $\delta m$ reflects the incertitude in the estimation of the length and the diameter of the nanotube obtained by atomic force microscopy (AFM). Alternatively, the mass resolution can also be determined from an Allan-like deviation of $f_0$ [19]; in so doing, we get $\delta m$ = 1.2 (±0.3) yg (section D of supplementary information).

Our ultra-sensitive resonator allows us to monitor adsorption of atoms and molecules with unparalleled accuracy. Upon dosing (i.e. sending) Xe atoms through a pinhole microdoser, the resonance frequency showed a tendency to decrease (Fig. 3a), indicating that Xe atoms were being adsorbed on the nanotube. On top of this trend, abrupt upward shifts in $f_0$ as high as ~5 MHz were also detected (red arrows). When we stopped dosing, no abrupt upward shift was observed (Fig. 3a). A possible origin of these upward shifts might be related to individual (or packets of) Xe atoms that either desorb into the vacuum or diffuse along the nanotube towards the clamping regions (Fig. 1a). The desorption-diffusion process might be aided by the impact of high energy Xe atoms which originate from a 300 K reservoir and whose energy follows the Boltzmann distribution (since the system is out of equilibrium in such a scenario, it is difficult to discriminate between desorption and diffusion events). We carried out 45 similar measurements with Xe and naphthalene, which all show a similar trend (section F of supplementary information).

We also observed downward shifts in $f_0$ that are consistent with single adsorption events. Figure 3b shows a series of such shifts obtained by dosing naphthalene ($C_{10}H_8$) molecules. The average frequency shift is $-3 \cdot 10^5$ Hz, and a shift occurs every 1-2 s. This is in agreement with the expected average frequency shift $-m_{C10H8}/2m_{NT} \cdot f_0 = -3.2 \cdot 10^5$ Hz due to the adsorption of a single $C_{10}H_8$ molecule ($m_{C10H8}$ is the mass of one $C_{10}H_8$ molecule and $m_{NT}$ the mass of the nanotube; see section C of supplementary information). Deviations from this value are attributed to the dependence of the frequency shift on the location of the adsorption along the nanotube axis (inset of Fig. 3b). As for the adsorption rate, we estimate it from the geometry of the chamber and the measured pressure (section A of supplementary information). An adsorption event is expected to occur on average every ~3 s, which is consistent with the measurements. Because of the predominant effect of desorption-diffusion, observing a series of frequency shifts as in Fig. 3b is rare. Future work will be devoted to engineering a single trapping site in the nanotube. We note that single adsorption events were previously detected for biomolecules and nanoparticles [2] that are at least two orders of magnitude heavier than $C_{10}H_8$ molecules. Since the interaction between these large objects and the resonator is strong, their immobilization is possible without trapping site.

Next we demonstrate that the adsorption of Xe atoms onto a nanotube surface is a thermally activated process. The lower inset of Fig. 4a shows the temperature dependence of $f_0$ while dosing Xe atoms onto the nanotube at a rate of ~22 atoms/s. At temperatures above $T$ ~58 K, the resonance frequency remains unchanged. Upon lowering the temperature, $f_0$ decreases. The number of Xe atoms per carbon atom is extracted using $N_{Xe}/N_C = m_C/m_{Xe} \cdot (f_{high}^2/f_0^2 - 1)$ with $f_{high}$ the resonance frequency at high temperature (59 K) and $m_C$ ($m_{Xe}$) the mass of a carbon (xenon) atom [5] ($f_0$ is essentially insensitive to the tension induced by Xe-Xe interaction that is two orders of magnitude weaker than the covalent C–C bonds [5]). Figure 4a shows that the temperature dependence of $N_{Xe}/N_C$ is consistent with a thermally activated behavior. We attribute this behavior to the balance of Xe atoms impinging on and departing from the nanotube: upon lowering $T$ the number of atoms on the nanotube increases as $N_{Xe} \propto \exp(E_b/k_B T)$ where $E_b$ is the binding energy (section G of supplementary information).

We extract the xenon-nanotube binding energy, $E_b = 131$ meV, from the slope in Fig. 4a. Measuring a second resonator yields $E_b = 110$ meV (section G of supplementary information). This energy is significantly lower than the xenon-graphite binding energy, which is 162 meV [20]. To understand this difference, we calculated the van der Waals interaction energy between a Xe atom and the different C atoms of a nanotube as a function of the xenon-nanotube separation (Fig. 4b) (section H of supplementary information). The resulting binding energy depends on the nanotube diameter $d$ (Fig. 4c). We get $E_b = 133$ meV for $d = 2.2$ nm, in agreement with the measurements. The binding energy is lower than the value reported for graphite, because (i) the nanotube curvature modifies the distance between the Xe atom and the C atoms (upper inset of Fig. 4a), and (ii) several graphene layers below the surface of graphite effectively contribute to the long-range van der Waals interaction.

The measurement in Fig. 4a illustrates an exciting new strategy to study adsorption, since it can be performed on an individual nanotube and with a resolution down to the atomic level (the error bar at low $N_{Xe}$ corresponds to 6 atoms). Gas adsorption on nanotubes has been the subject of intense efforts in view of possible gas storage applications [21-25]. However, previous adsorption measurements were notoriously difficult to interpret, since they were carried out on films of nanotubes where atoms can diffuse through the film and where the binding energy depends on the different adsorption sites. We avoid these limitations by measuring a single nanotube.

In conclusion, the exquisite mass resolution of nanotube resonators holds promise for various scientific and technological applications. We have demonstrated adsorption measurements on an individual nanotube and with a resolution down to the atomic level. The frequency stability of nanotube resonators may enable high-sensitive magnetometry measurements of a magnetic nano-object attached to the nanotube [26]. A new generation of mass spectrometers based on nanotube resonators might be developed that identify the chemical nature of individual atoms and molecules. For this, the atoms and molecules to be weighed could be trapped at one specific site of the nanotube. An approach to fabricating a single trapping site is to locally modify the carbon lattice by means of electrochemistry [27]. The trapping site would (i) allow a direct measure of the mass, which is not affected by the random location of the atoms and molecules to be weighed, and (ii) reduce fluctuations in $f_0$ induced by desorption-diffusion.

**Acknowledgements**

We acknowledge support from the European Union through the RODIN-FP7 project, the ERC-carbonNEMS project, and a Marie Curie grant (271938), the Spanish ministry (FIS2009-11284, TEC2009-06986, FIS2009-12721-C04-03, CSD2007-00041), and the Catalan government (AGAUR, SGR). We thank Brian Thibeault (Santa Barbara) for help in fabrication.

**Author Contributions**

J.C. fabricated the devices, developed the measurement setup and performed the measurements. R. R. carried out the calculations of the binding energy. A.B. supervised the work. All authors contributed to discussing the results and writing the manuscript.

**Author information**

Correspondence and requests for materials should be addressed to A.B. (adrian.bachtold@cin2.es).

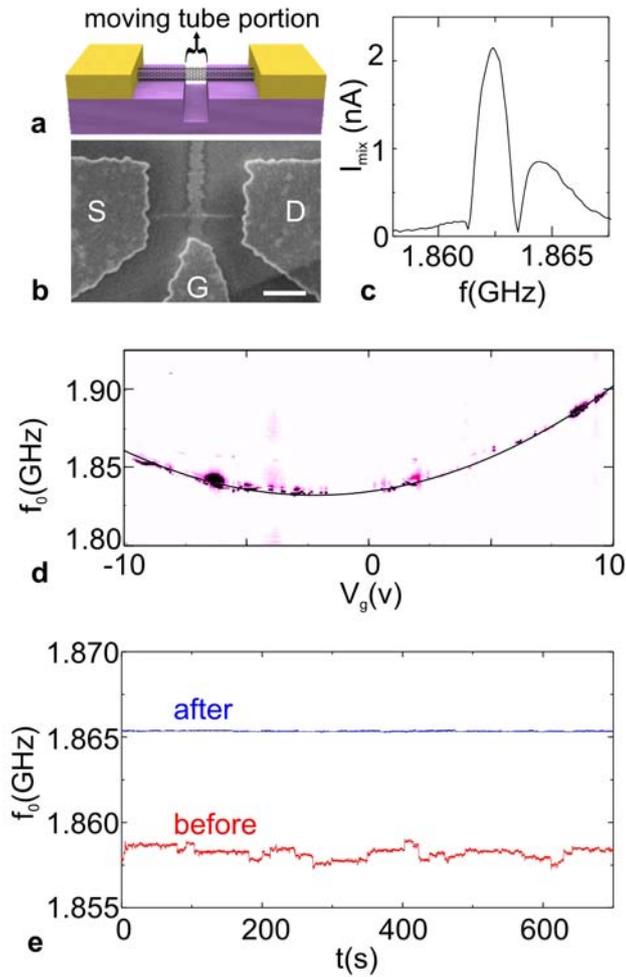

**Fig.1 Device and characterization. a,b,** Schematic and scanning electron microscopy image of the device. The diameter of the nanotube is $d = 1.7$ nm and the length of the suspended nanotube section is $L_{tube} \approx 150$ nm (as measured with AFM). Scale bar: 300 nm. **c,** Mechanical resonance obtained by measuring the mixing current $I_{mix}$ as a function of the driving frequency $f$ using the FM mixing technique at 6 K. The amplitude of the applied FM voltage is 4 mV and the integration time of the lock-in amplifier used to measure $I_{mix}$ is 30 ms. The resonance is asymmetric (the lobe on the right side of the central peak is larger than the lobe on the left side); this asymmetry is attributed to the Duffing force, since it is less pronounced at lower driving force [15]. Hysteresis jumps are not observed even at higher driving force [15]. This resonance

likely corresponds to the fundamental eigenmode, since it is the resonance with the lowest $f_0$ and the highest $I_{mix}$ (we detect a second resonance at 3.4 GHz). **d,** Resonance frequency as a function of the voltage applied on the side-gate electrode at 6 K (obtained by measuring $I_{mix}$ as a function of $f$ and $V_g$). The intensity of $I_{mix}$ varies with $V_g$ because of the variation of the transconductance. **e,** Resonance frequency as a function of time at 6 K before and after current annealing.

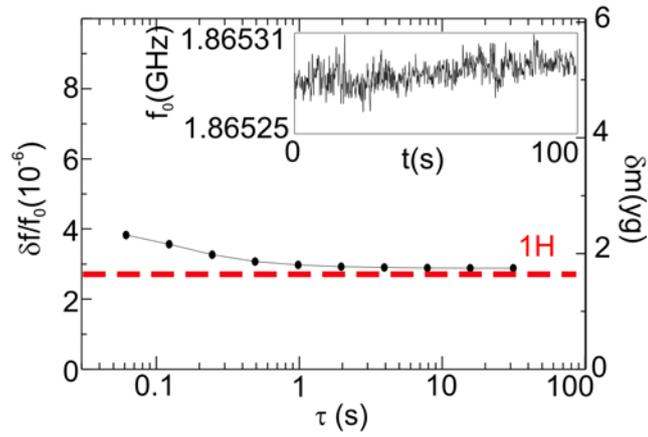

**Fig.2 Standard error of the resonance frequency and corresponding mass resolution as a function of averaging time at 5.5 K**. The red dashed line corresponds to the mass of one hydrogen atom. The inset shows the resonance frequency as a function of time. The amplitude of the applied FM voltage is 4 mV.

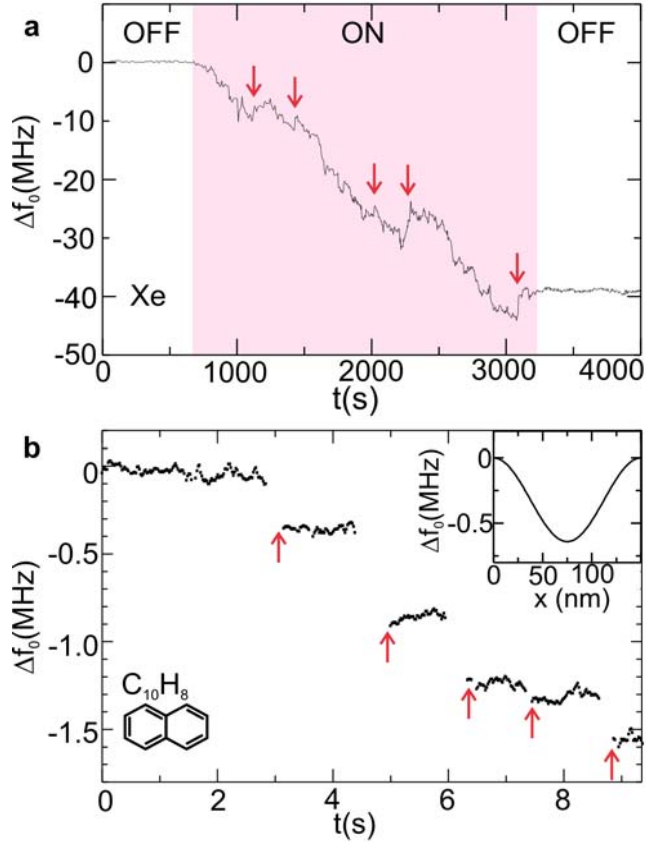

**Fig.3 Adsorption of Xe atoms and naphthalene molecules. a,** Resonance frequency as a function of time at 6 K as xenon atoms are being dosed onto the nanotube. Red arrows indicate some of the abrupt upward shifts discussed in the text. The shaded area corresponds to the time when Xe atoms are dosed; Xe atoms arrive directly from the microdoser onto the nanotube so the dosing rate cannot be measured. The resonance frequency is obtained by continuously measuring $I_{mix}$ as a function of $f$. **b,** Resonance frequency as a function of time at 4.3 K when naphthalene molecules are being dosed. Red arrows point to the shifts of the resonance frequency consistent with the adsorption of $C_{10}H_8$ molecules. The resonance frequency is measured using a computer-controlled feedback loop. The inset shows the expected shift in $f_0$ as a function of the position of the $C_{10}H_8$ adsorption along the 150 nm long nanotube. Xe atoms and $C_{10}H_8$ molecules are admitted from a gas reservoir at 300 K into the vacuum chamber through a pinhole

microdoser. In the case of $C_{10}H_8$ dosing, unwanted contaminants are removed by freeze-pump cycles of the gas reservoir prior to adsorption experiments [28].

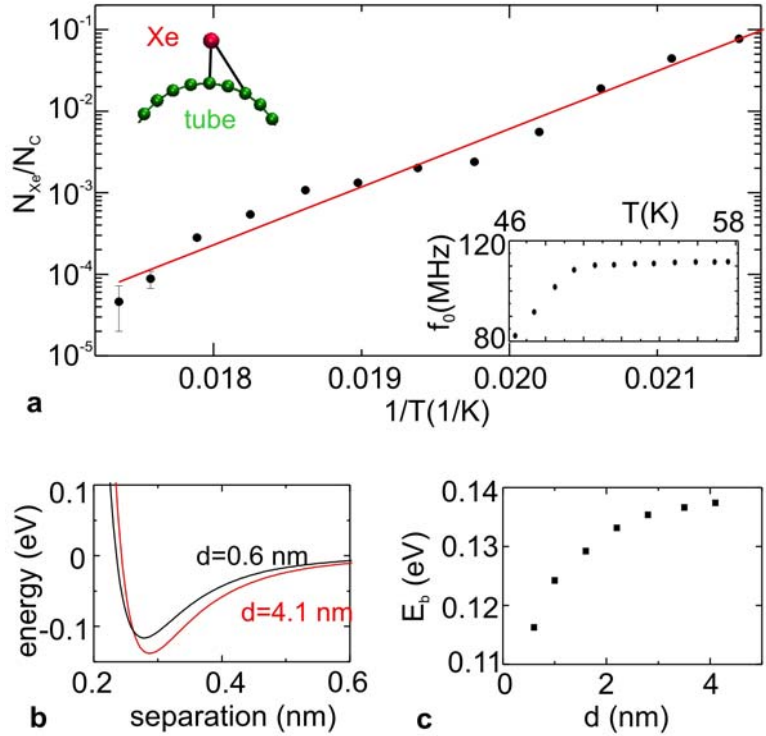

**Fig.4 Binding energy between a Xe atom and a nanotube. a,** Number of Xe atoms per C atom as a function of $1/T$ when dosing Xe atoms onto the nanotube with a rate of ~22 atoms/s. The lower inset shows the resonance frequency as a function of temperature. The device is different from the one of Fig. 1-3; the nanotube was grown in the last fabrication step over a predefined trench separating two electrodes [15,29]. The diameter of the nanotube is $d = 2.2$ nm and the length of the suspended nanotube section is $L_{tube}$= 2200 nm (as measured with AFM on the nanotube section lying on the electrodes). The upper inset shows a Xe atom interacting with the C atoms of a nanotube. Note that the temperature dependence of $N_{Xe}/N_C$ is expected to follow a thermal activation behavior over a limited temperature range; $N_{Xe}/N_C = 10^{-5}$ corresponds to about one Xe atom and a monolayer of Xe atoms forms at $N_{Xe}/N_C = 0.1 - 1$. **b,** Calculated interaction energy between a Xe atom and a nanotube as a function of their separation for two different nanotube diameters. **c,** Calculated binding energy as a function of nanotube diameter.